\documentclass[aps,prb,twocolumn,showpacs, citeautoscript]{revtex4-1}

\usepackage{amsmath}
\usepackage{amssymb}
\usepackage{graphicx}
\usepackage{bm}

\usepackage{color}
\usepackage[colorlinks, linkcolor=blue, citecolor=blue, urlcolor=blue,
breaklinks]{hyperref}

\begin{document}

\title{Simulation of supersymmetric quantum mechanics in a Cooper-pair box
shunted by a Josephson rhombus}
\author{Jascha Ulrich}
\author{Daniel Otten}
\author{Fabian Hassler}
\affiliation{JARA-Institute for Quantum Information, RWTH Aachen, D-52074,
  Germany}
\date{November 2015}

\pacs{%
  03.67.Ac, 
  11.30.Pb, 
  85.25.Cp, 
  42.50.Pq 
}

\begin{abstract}
Supersymmetries in quantum mechanics offer a way to obtain degeneracies in the
excitation spectrum which do not originate from selection rules. The mechanism
behind the degeneracies is the same as the one that leads to the miraculous
cancellations of divergences in supersymmetric field theories found in the
high energy physics context.  There is up to now no realistic proposal of
non-integrable systems that show level degeneracies due to a supersymmetric
structure.  Here, we propose an implementation of a quantum-mechanical
supersymmetry in a Cooper-pair box shunted by a Josephson junction rhombus
which is effectively $\pi$-periodic in the superconducting phase difference.
For a characteristic ratio between the strength of the $2\pi$- and the
$\pi$-periodic junction, we find a two-fold degeneracy of all the energy
levels all the way from the weak junction/charge qubit limit to the strong
junction/transmon regime.  We provide explicit experimental values for the
parameters of the system and show that tuning in and out of the supersymmetric
point is easily achieved by varying an external gate voltage.  We furthermore
discuss a microwave experiment to detect the supersymmetry and conclude that
it can indeed be implemented with currently existing Josephson junction
technology.
\end{abstract}

\maketitle

The macroscopic quantum mechanics of superconducting circuits has allowed the
experimental simulation of many complex quantum phenomena such as phase
transitions \cite{oudenaarden:96}, quantum spins \cite{neeley:09}, or dynamics
in open systems \cite{li:13}.  Theoretically, the quantum simulation of
intricate subjects such as Hawking radiation~\cite{nation:09} and lattice
gauge theories \cite{doucot:04, marcos:13, heck:14} has been proposed.  In the
plethora of phenomena that can be simulated with the help of superconducting
circuits~\cite{georgescu:14, paraoanu:14}, degeneracies due to
quantum-mechanical supersymmetries have notably been absent.  Typically,
degeneracies in the spectrum arise when the Hamiltonian commutes with all
group elements of a non-Abelian symmetry which translates into selection rules
demanding vanishing off-diagonal and equal diagonal matrix elements of the
Hamiltonian within the same irreducible representation \cite{messiah}.  The
degeneracy of the states thus always follows from the dimension of the
representation.  Supersymmetry on the other hand does not simply forbid
different states to couple but it makes sure that in each order of
perturbation theory there is always a pair of terms canceling each other
\cite{witten:81,ulrich:14}.

It is intriguing that the degeneracies of supersymmetric quantum
mechanics occur by the same mechanism \cite{witten:81} that leads to a
miraculous cancellation of divergences in supersymmetric field theories and
makes supersymmetries an important tool of particle physics~\cite{aitchison}.
In trivial cases like the free particle~\cite{rau:04} or the Jaynes-Cummings
model~\cite{andreev:89}, however, the supersymmetric structure is irrelevant
since the spectrum is exactly solvable.  In order to deepen the connection to
the ideas in the high-energy context, it is thus of vital importance to
propose a non-integrable system where the level degeneracy can be solely
understood by its supersymmetric structure.

\begin{figure}[tbp]
\includegraphics[width=0.7\columnwidth]{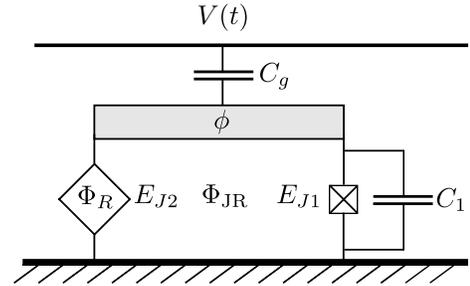}
\caption{Setup simulating the Hamiltonian Eq.~\eqref{eq:ham_system_eff}.
A Josephson junction (crossed box) with Josephson energy $E_{J1}$ and
capacitance $C_1$ couples the Cooper-pair box, a superconducting island with
superconducting phase $\phi$, to a ground superconductor (diagonally striped)
with phase $\phi_0 = 0$.  A Josephson rhombus (depicted as a rhombus) provides
an additional shunt to the ground which generates a $\cos(2\phi)$ Josephson
coupling of strength $E_{J2}$ when the rhombus is threaded by a flux $\Phi_R =
\Phi_0/2$.  We assume that an additional flux $\Phi_{JR}=-\Phi_R/2$ threads
the loop between the standard junction and the rhombus, which requires a
second flux line that can be controlled separately.  A capacitance $C_g$
couples the system to a transmission line at the voltage
$V(t)$.}\label{fig:setup}
\end{figure}

In this paper, we show that shunting a Cooper-pair box with a Josephson
junction rhombus simulates a highly nontrivial supersymmetry that can be
experimentally realized with today's Josephson junction technology.  This
proposal combines ideas of implementing supersymmetry in purely bosonic
systems~\cite{plyushchay:96} with the quantum-mechanical supersymmetry that
has recently been proposed for superconductors hosting fermionic Majorana
bound states~\cite{ulrich:14}.  The only nonstandard component of our proposal
is the Josephson rhombus.  The Josephson rhombus is a junction between two
superconductors that allows only tunneling of pairs of Cooper pairs.
Consequently, its current-phase relation is $\pi$-periodic~\cite{doucot:02}.
Josehson rhombi have previously been proposed as building blocks for
topologically protected qubits~\cite{ioffe:02, doucot:12} that have been shown
to be experimentally feasible~\cite{gladchenko:09}.  Additionally, they have
been employed for the experimental realization of qubits based on the
Cooper-pair parity~\cite{bell:14}.  Up to now, theoretical studies on the
Josephson rhombi have mainly been focused on the semi-classical regime
\cite{ioffe:02a, protopopov:04}.  For our setup, we study a Josephson rhombus
in the charging limit where it is adiabatically coupled to the superconducting
island and generates a $\pi$-periodic Josephson coupling of a specific
$\cos(2\phi)$ form.  A supersymmetry is then obtained for a characteristic
ratio between the strength of the conventional and the $\pi$-periodic
Josephson junction, see below.

Our system of interest is depicted in Fig.~\ref{fig:setup}.  It is an
extension of the conventional Cooper-pair box~\cite{bouchiat:98, nakamura:99},
which consists of a superconducting island with superconducting phase $\phi$
and a ground superconductor with phase $\phi_0 = 0$ which are coupled by a
Josephson junction with Josephson energy $E_{J1}$ and capacitance $C_1$.  A
capacitance $C_g$ couples the system to a transmission line biased at a DC
voltage $V(t) = V_g$.  We add an additional shunt to the ground through a
Josephson rhombus with capacitance $C_\diamond$, which, as we will discuss in
more detail below, generates a $\pi$-periodic Josephson energy $-E_{J2}
\cos(2\phi)$ when threaded by a flux $\Phi_R = \Phi_0/2$, where $\Phi_0 =
h/2e$ is the superconducting flux quantum and $E_{J2}$ the effective junction
energy.  Taking into account an additional flux $\Phi_{JR} = -\Phi_R/2$ in the
loop between the conventional junction and the rhombus, we obtain the
effective low-energy Hamiltonian
\begin{align}\label{eq:ham_system_eff}
H_\text{eff} = 4 E_{C_\Sigma} (n - n_g)^2 - E_{J1} \cos \phi - E_{J2}
\cos(2\phi),
\end{align}
where $n = -i \partial/\partial \phi$ is the number of Cooper pairs on the
 island, $E_{C_\Sigma} =
e^2/2 C_\Sigma$ with $C_\Sigma = C_1 + C_g + C_\diamond$ is the total charging
energy of the island, and $n_g = C_g V_g/2e$ is the induced offset charge in
units of $2e$.  The Hamiltonian \eqref{eq:ham_system_eff} does not admit an
analytic solution.  Its only symmetry is the operation $K: \phi \mapsto -\phi$
at the point $n_g=0$ which due to its Abelian nature does not lead to any
degeneracy.  However, as we will show below, for a specific ratio of the
energy scales all excited levels are degenerate due to a supersymmetry.

In the simplest setting, a Hamiltonian $H_Q$ is called supersymmetric when
there exists a Hermitian involution $K$ with $K^2 = 1$ that commutes with
$H_Q$ and a Hermitian supercharge $Q$ which anticommutes with $K$ and
factorizes the Hamiltonian $H_Q = Q^2$~\cite{cooper, combescure:04}.  The
sectors of $H_Q$ are then characterized by $K$ according to $H_Q = P_+ H_Q P_+
+ P_- H_Q P_-$ with $P_\pm = (1 \pm K)/2$ and are intertwined through the
relation $P_\pm Q = Q P_\mp$ which guarantees that to each eigenstate
$|a\rangle$ to energy $E_a > 0$ in one sector there is a partner state
$(Q/\sqrt{E_a})|a\rangle$ to the same energy in the other sector.  To see how
this relates to our system, let us introduce the supercharge $Q$ and the
involution $K$ according to~\cite{ulrich:14}
\begin{align}\label{eq:supercharge}
Q = 2 \sqrt{E_{C_\Sigma}} \bigl( n + i \alpha \sin \phi\bigr)(-1)^n, \quad
K \phi K = -\phi,
\end{align}
where $\alpha$ is a free parameter.  Note that the supercharge $Q$ is
Hermitian since the addition/removal $e^{\pm i \phi}$ of a Cooper-pair
anticommutes with the Cooper-pair parity ${(-1)}^n$ of the island.  We then
find that with $\alpha = \sqrt{E_{J2}/2 E_{C_\Sigma}}$ and up to irrelevant
constants, the supercharge $Q$ squares to the Hamiltonian
Eq.~\eqref{eq:ham_system_eff} at the point $n_g = 0$ and
\begin{align}\label{eq:susy_params}
E_{J1} &= \sqrt{8 E_{J2} E_{C_\Sigma}},
\end{align}
where the system Eq.~\eqref{eq:ham_system_eff} is supersymmetric.  The exotic
feature of the supersymmetry to note here is the preservation of the
degeneracy of the energy levels all the way from the charge qubit regime with
$\alpha \ll 1$ to the transmon regime $\alpha \gg 1$ as long as
\eqref{eq:susy_params} is fulfilled.  The ``hidden'' character of this
degeneracy is underlined by the highly nonlocal form of the
supercharge~\eqref{eq:supercharge}.

\begin{figure}[tpb]
\includegraphics[width=.95\columnwidth]{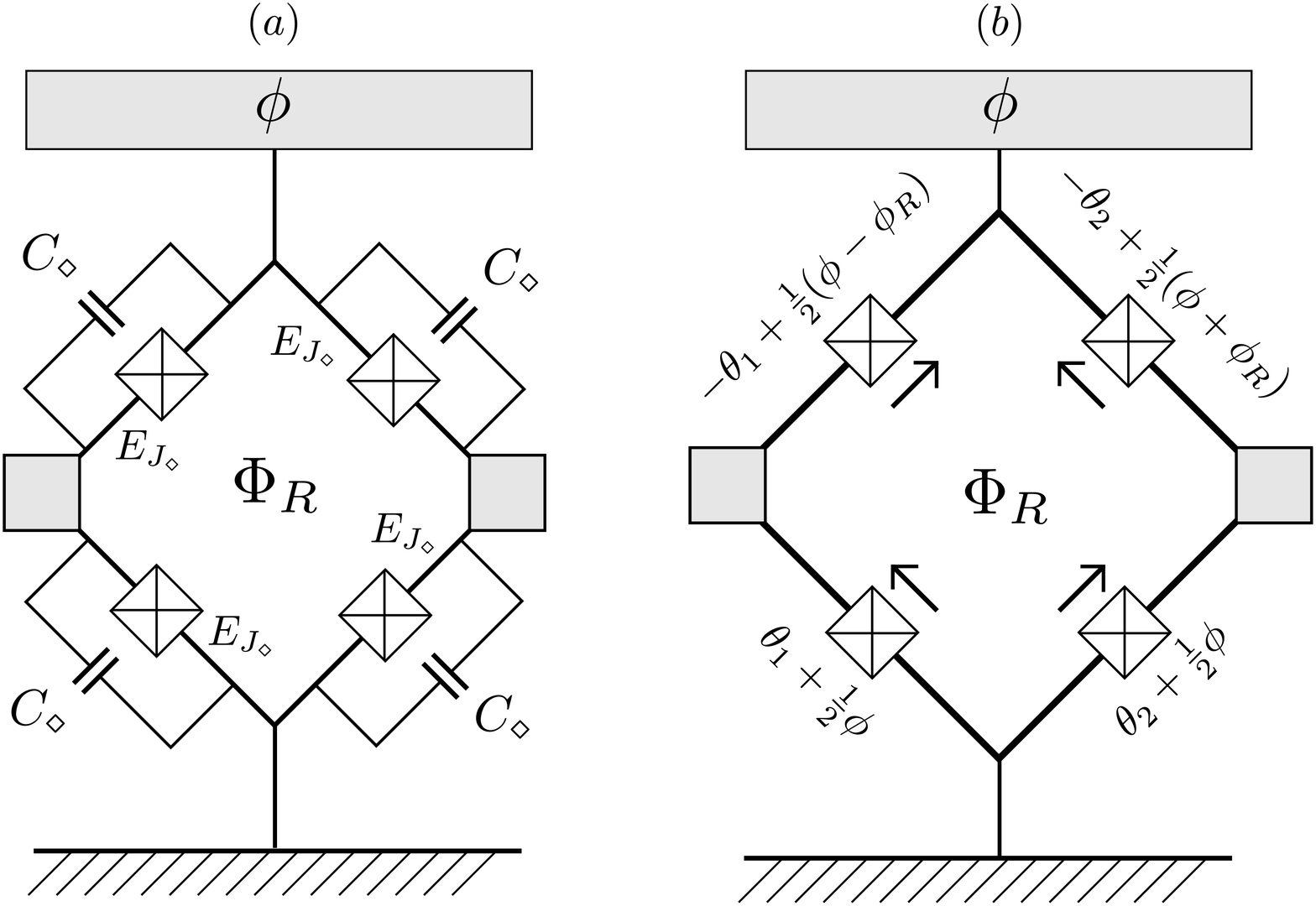}
\caption{%
($a$) Circuit of the Josephson rhombus  consisting of a loop interrupted by
four Josephson junctions with Josephson energies $E_{J_\diamond}$  and
capacitances $C_\diamond$.  When the loop is threaded by a flux of
$\Phi_R=\Phi_0/2$, the transport of single Cooper-pairs through the device is
suppressed by  destructive interference between the tunneling events through
the left and the right arm of the rhombus, rendering the rhombus eigenenergies
$\pi$-periodic in the fixed phase difference $\phi$ between the top island and
the ground superconductor.  ($b$)  Choice of the gauge-invariant phase
differences across the links with the arrows indicating their orientation.  The
phases $\theta_j$ are conjugate to the charges $n_j = -i
\partial/\partial{\theta_j}$ of the superconducting islands (gray squares) in
the two arms of the rhombus. The phases  add up to the reduced flux
$\phi_R = 2\pi\Phi_R/\Phi_0$.}\label{fig:rhombus}
\end{figure}
\begin{figure*} \includegraphics[width=1.\textwidth]{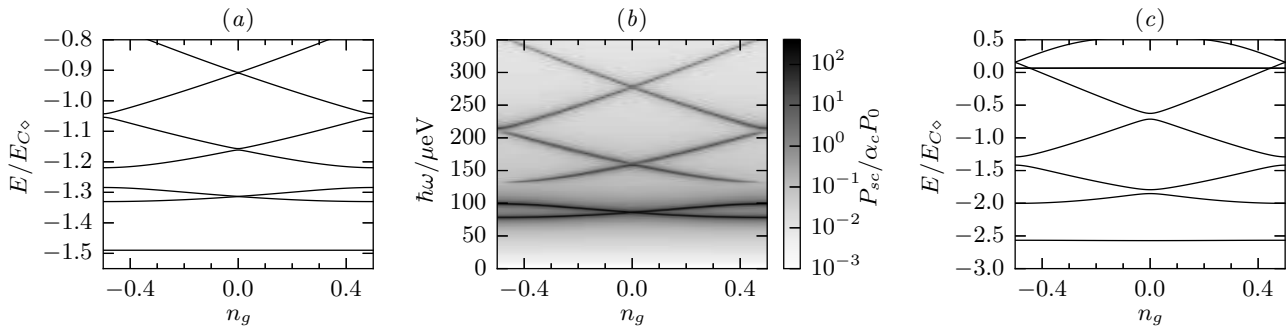}
\caption{ 
Energy spectra of the full Hamiltonian Eq.~\eqref{eq:ham_system_full} as a
function of the dimensionless offset charge $n_g = C_g V_g/2e$ for $\alpha=2,
\eta=0.7$ in ($a$) and $\alpha=1, \eta=1$ in ($c$).  For a given
$\eta=E_{J\diamond}/2 E_{C\diamond}$, we numerically determine the strength
$E_{J2}$ of the $\pi$-periodic component of the ground state energy
$\epsilon_0(\phi)$ of the rhombus.  The value of the charging energy is then
obtained from the relation $E_{C_\Sigma} = E_{J2}/2 \alpha^2$.  The strength
of the Josephson coupling $E_{J1}$ is fixed at a value $ 4 E_{C_\Sigma}
\alpha$ which corresponds to the supersymmetric point,
cf.~\eqref{eq:susy_params}.  It can be seen in ($a$) that all the excited
levels cross at $n_g=0$, confirming the validity of the effective
supersymmetric model~\eqref{eq:ham_system_eff} beyond the perturbative regime
$\eta \ll 1$.  As shown in ($c$) the former supersymmetric level crossings at
$n_g = 0$ turn into avoided crossings at a slightly elevated $\eta$.  This
signals the breakdown of the rhombus/island decoupling and thus restricts the
mapping of the full Hamiltonian to the supersymmetric model.  Additionally,
the first excited rhombus level can be seen as a horizontal line around
$E/E_{C\diamond} = 0$.
($b$) Plot of the power loss $P_\text{sc}$ at frequency $\omega$ of the
transmission line coupled to the system with parameters given in ($a$) which
corresponds to the experimental parameters $(C_1 + C_g)/C_\diamond = 80$,
$E_{J1} = 0.1 \, E_{C\diamond}$, and $E_{C\diamond} = 500\,\mathrm{\mu eV}$,
cf.  \eqref{eq:power_loss}.  The power loss is measured in units of the
injected intensity $P_0$ and a dimensionless coupling constant $\alpha_C$.
Here, we assume that the system once exited relaxes fast with a rate $\Gamma =
0.5\,\mathrm{\mu eV}/\hbar$ into degrees of freedom different from the
transmission line such that the resonance condition is indicated by a dip of
size $P_\text{sc}$ in the transmitted intensity.  Since the ground state is
nearly insensitive to changes in $n_g$, the power loss can be directly
compared to the spectrum shown in ($a$).
}\label{fig_spectra} 
\end{figure*}

From the above, we see that an (effective) Josephson junction with a
$\pi$-periodic Josephson energy of the form $-E_{J2} \cos(2\phi)$ is crucial
for supersymmetry.  Such a circuit element is provided by the Josephson
rhombus shown in Fig.~\ref{fig:rhombus}($a$) \cite{doucot:02}.  It is a two
arm Cooper-pair interferometer connecting the superconducting island to the
ground in which single Cooper-pairs tunneling through the left and right arm
of the rhombus interfere destructively due to a magnetic flux $\Phi_R =
\Phi_0/2$.  Each arm contains two Josephson junctions connected in series with
Josephson energy $E_{C\diamond}$ and capacitance $C_\diamond$.  We show that
in the charging regime $\eta = E_{J\diamond}/2 E_{C\diamond} \lesssim 1$, the
ground state energy of the rhombus is well approximated by $\epsilon_0(\phi)
\approx -E_{J2} \cos(2\phi)$, where $\phi$ is the (fixed) phase difference
between island and ground.  Furthermore, we argue that the weak coupling to
the island permits an adiabatic decoupling leading to the effective
Hamiltonian~\eqref{eq:ham_system_eff}.  To this end, let us denote the
Cooper-pair number of the superconducting islands in the left and right arm by
$n_1$, $n_2$ and choose the gauge-invariant phase differences across the
junctions as indicated in Fig.~\ref{fig:rhombus}($b$).  Taking into account
the additional flux $\Phi_{JR} = -\Phi_0/4$ from Fig.~\ref{fig:setup}, the
Hamiltonian of the full system assumes the form
\begin{align}\label{eq:ham_system_full}
H = 4 E_{C_\Sigma} (n - n_g)^2 - E_{J1} \cos \phi + H_\diamond,
\end{align}
which corresponds to the effective model Eq.~\eqref{eq:ham_system_eff} with
$-E_{J2}\cos(2\phi)$ replaced by the rhombus Hamiltonian
\begin{align}
H_\diamond=2E_{C\diamond}(n_1^2+n_2^2)+V_\diamond(\theta_1,\theta_2,\phi).
\end{align} 
Here, $E_{C\diamond} = e^2/2 C_\diamond$ is a charging energy, the phases
$\theta_1$, $\theta_2$ are conjugate to $n_1 = -i \partial/\partial \theta_1$,
$n_2 = -i \partial/\partial \theta_2$, and the potential $V_\diamond$ reads
\begin{align*}
V_\diamond = -E_{J\diamond} \sum_{j=1,2} \Bigl[\cos(\tfrac{\phi}{2}+\theta_j)
+\cos\bigl(\tfrac\phi2 + (-1)^j \tfrac{\phi_R}2 -\theta_j\bigr) \Bigr],
\end{align*}
where $\phi_R = 2\pi\Phi_R/\Phi_0$ is the reduced flux.  For fixed $\phi$, the
Hamiltonian $H_\diamond$ possesses instantaneous eigenstates $|n; \phi\rangle$
with eigenvalues $\epsilon_n(\phi)$.  The destructive interference of single
Cooper-pair tunneling is expressed by the fact that exchanging the two
tunneling paths and advancing $\phi$ by $\pi$ is a symmetry of the
system%
\,\cite{Note1},
%
%
demanding $\pi$-periodicity of $\epsilon_n(\phi)$.  At half a flux quantum,
time-reversal is an additional symmetry demanding an even $\epsilon_n(\phi)$.
Consequently, the ground state energy $\epsilon_0(\phi)$ must be of the form
$\epsilon_0(\phi) = -\sum_n E_{J2n} \cos(2n\phi)$.  By perturbation theory in
$\eta$, we find that the desired $\pi$-periodic component
$E_{J2}/E_{C\diamond} = 7 \eta^4/4$ dominates, with $E_{J4}/E_{C\diamond} =
-68687 \eta^8/36864$ and $E_{J2m}/E_{C\diamond} \propto \eta^{4m}$.  In the
following, we are interested in the regime $\eta \lesssim 1$ and thus we
determine $E_{J2}$, $E_{J4}$ numerically from $\epsilon_0(\phi)$ whenever
needed.  We find that $E_{J2}$ stays at least an order of magnitude larger
than $E_{J4}$ up to $\eta \approx 1$.  Since the above form of the rhombus
energies is due to symmetry, differing Josephson couplings $E_{J\diamond, L/R}
= E_{J\diamond}(1\pm \delta/2)$ in the left and right rhombus arm will in
general induce a finite $2\pi$-periodic Josephson coupling whose strength
scales perturbatively as $4 \eta^2 \delta \,E_{C\diamond}$.  Comparison with
$E_{J2}$ yields that the effects of asymmetry are negligible for $\eta^2 \gg
\delta$.

Projecting the Hamiltonian Eq.~\eqref{eq:ham_system_full} onto the
instantaneous rhombus ground state $|0; \phi \rangle$ and using
$\epsilon_0(\phi) \approx -E_{J2} \cos(2\phi)$ leads by standard
methods~\cite{bohm} to the effective Hamiltonian $H_\text{ad} = H_\text{eff} +
4 E_{C_\Sigma} \sum_{n > 0} |A_{n0}|^2$, where $A_{nm} = i \langle n; \phi |
\partial_\phi | m; \phi \rangle$ is the induced vector potential describing
the nonadiabatic corrections.  In deriving $H_\text{ad}$, we have used that
the term $A_{00}$ vanishes since the states $|m; \phi\rangle$ can be chosen
real.  Due to time-reversal and rhombus symmetry, $|A_{n0}|$ is even in $\phi$
and thus does not couple the supersymmetric partners at $n_g = 0$.  Since the
gap to the next pair of supersymmetric states is at least of order
$E_{C_\Sigma}$, the effects of $|A_{n0}|^2   =$ $ |\langle n; \phi | \partial_\phi
H_\diamond | 0 ; \phi\rangle|^2/[\epsilon_n(\phi) - \epsilon_0(\phi)]^2 $ $
\propto \eta^2$ are negligible for $\eta \ll 1$.  The coupling to the excited
rhombus levels that we do not take into account in the projected Hamiltonian
$H_\text{ad}$ is suppressed by even higher orders in $\eta$.

The supersymmetry becomes trivial for $\alpha \rightarrow 0$ where one
recovers the supersymmetry of the free particle~\cite{rau:04}.  We therefore
aim for the most interesting regime of $\alpha \approx 1$, where all the terms
in the Hamiltonian \eqref{eq:ham_system_eff} are of the same order.  While the
adiabatic decoupling of the rhombus is most robust for large scale separation
$\eta \ll 1$ between the rhombus and the island, our perturbative results for
$E_{J2}$ show that this also implies $\alpha \propto \eta^2 \ll 1$.  The
regime $\alpha \approx 1$ thus requires moderately large $\eta$ for which we
numerically show that the adiabatic decoupling still works.
Figure~\ref{fig_spectra}($a$) shows the numerical results for the spectrum of
the full Hamiltonian as a function of the offset charge $n_g$ for $\alpha = 2$
and $\eta = 0.7$.  We highlight that \emph{all} excited levels, including the
levels higher in energy not visible in Fig.~\ref{fig_spectra}($a$), become
doubly degenerate as $n_g$ approaches zero.  This degeneracy of all excited
states in complete absence of selection rules gives a clear signature of
supersymmetry.  Figure~\ref{fig_spectra}($c$) shows the high sensitivity of
the supersymmetry to the choice of $\eta$.  For $\alpha = 1$ and $\eta = 1$,
supersymmetry at $n_g = 0$ is clearly destroyed by non-adiabatic corrections
in the Hamiltonian $H_\text{ad}$.  Moreover, the first excited level of the
rhombus showing up as a horizontal line in the upper region of
Fig.~\ref{fig_spectra}($c$) is visible.

The spectrum of the system can be read out with the help of the transmission
line coupled to the island by injecting a voltage $V(t) = V_g + V_\omega
\cos(\omega t)$ with the AC amplitude $V_\omega$ at the frequency $\omega$.  For
a transmission line characterized by the admittance $Y_\text{tl}$, this
corresponds to an average injected intensity $P_0 = \tfrac{1}{2} Y_\text{tl}
V_\omega^2$.  According to the Hamiltonian~\eqref{eq:ham_system_full}, the AC
voltage drives transitions in the system through the coupling $H_C = g_C
V(t) n$ with $g_C = 2e C_g/C_\Sigma$.  Due to the coupling to the charge $n$,
the driving is most effective for $\alpha \gtrsim 1$.  We assume that the
system relaxes dominantly into channels different from the transmission line
with a rate $\Gamma \gg \Gamma_\text{abs}$, where $\Gamma_\text{abs}$ is the
rate of photon absorption.  Measuring in transmission, the absorption of photons
is then signaled by a reduced transmitted intensity $P_t$ with respect to the
incoming intensity $P_0$.  With $\Gamma \gg \Gamma_\text{abs}$, photons of
energy $\hbar \omega$ exclusively drive transitions from the ground state to
excited states and the scattered intensity $P_\text{sc} = P_0 - P_t$ follows
as $P_\text{sc} = \hbar \omega \Gamma_\text{abs}$.  For the calculation of
$\Gamma_\text{abs}$, we employ Fermi's golden rule.  We find
\begin{align}\label{eq:power_loss}
P_\text{sc}&= 4 \alpha_C P_0  
\sum_{n>0} 
 \frac{\hbar^2 \omega\Gamma 
  |\langle n | i \partial_\phi |
  0 \rangle |^2}
  {(E_n -E_0 - \hbar \omega)^2 + \hbar^2\Gamma^2},
\end{align}
denoting the eigenstates to energy $E_n$ of the full
Hamiltonian~\eqref{eq:ham_system_full} by $|n\rangle$; here, the dimensionless
constant $\alpha_C$ is given by $\alpha_C = g_C^2/\hbar Y_\text{tl} = 2\pi
C_g^2/C_\Sigma^2 Z_0 Y_\text{tl}$ with the superconducting impedance quantum
$Z_0 = h/4e^2 \approx 1\,\mathrm{k\Omega}$.  The condition $\Gamma_\text{abs}
\ll \Gamma$ translates into $g_C V_\omega \ll \hbar \omega$.  The validity of
Fermi's golden rule for the rate calculation on the other hand demands
$\alpha_C \omega \ll \Gamma$.

The system with $\alpha = 2$ presented in Fig.~\ref{fig_spectra}($a$)
corresponds to the experimental parameters $\eta = 0.7$, $(C_1 +
C_g)/C_\diamond = 80$ and $E_{J1} = 0.1 E_{C\diamond}$.  In
Fig.~\ref{fig_spectra}($b$), we show the numerical results for the
scattered intensity $P_\text{sc}$ as a function of the offset charge $n_g$ and
the radiation frequency $\omega$ with the system parameters stated above.  We
assume the experimental scale $E_{C\diamond} = 500\,\mathrm{\mu eV}$.  As
visible from Fig.~\ref{fig_spectra}($a$), the ground state is almost
insensitive to changes of $n_g$ at $\alpha = 2$ such that the scattered
intensity corresponds directly to the energy spectrum of the system.  The
scattered intensity is strongest for the first degenerate pair of levels which
reflects the fact that states higher in energy show a behavior closer to the
charging regime than the low-energy states. Note that the whole spectrum and
the level crossings of all excited states at the supersymmetric point
\eqref{eq:susy_params} with $n_g = 0$ can be clearly observed.

As a last point, let us comment on the susceptibility to imperfections in
design.  The above analysis was based on the rhombus symmetry which is
violated both by stray offset charges or parameter variations within the
offset arms as parametrized by $\delta$.  As we explain in more detail in the
Appendix, the susceptibility to stray offset charges can in practice be
reduced by adding inductive shunts to the ground within the rhombus
arms~\cite{bell:14, doucot:12}.  For the system parameters chosen above,
numerical checks show that symmetry violations $\delta$ are tolerable up to a
few percent.  The most demanding experimental requirements are thus the
reproducible parameter of the rhombus junctions.  In contrast, deviations in
$C_1/C_\diamond$ can be accounted for by the Josephson coupling $E_{J1}$ which
can be tuned easily.

To conclude, we have shown that a quantum-mechanical supersymmetry arises in a
Cooper-pair box when it is shunted by a Josephson rhombus operated in the
charging regime, where it is weakly coupled to the Cooper-pair box and
generates an effective $\pi$-periodic Josephson coupling of the form
$\cos(2\phi)$.  The supersymmetry is nontrivial since there are no selection
rules and an analytic solution is impossible, but yet, it guarantees an exact
degeneracy of all excited levels.  We have shown that the supersymmetry can be
detected through microwave spectroscopy and tuning in and out of the
supersymmetric point is easily possible by tuning the gate voltage.  We have
proposed realistic device parameters, paving the way for an experimental
exploration of the exotic degeneracies brought by supersymmetries in the near
future.

The authors acknowledge financial support from the Alexander von Humboldt
Foundation and the Deutsche Forschungsgemeinschaft under Grant No.\,HA
7084/2-1 as well as useful discussions with Christoph Ohm.

\appendix
\section{Device with inductive shunts}
\label{sec:appendix_inductance}
\begin{figure}[tpb]
\includegraphics[width=.95\columnwidth]{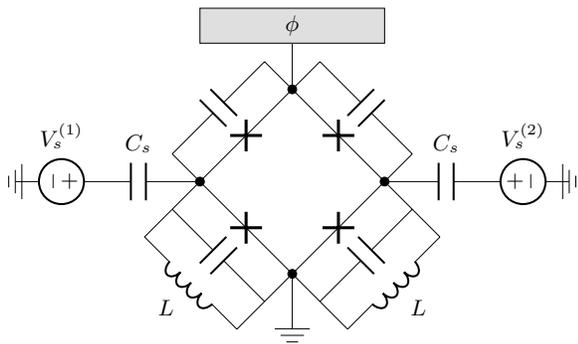}
\caption{%
Rhombus design as proposed theoretically in Ref.~\onlinecite{doucot:12} which
includes inductances $L$ that shunt the islands in the rhombus arms and reduce
the susceptibility to stray offset charges coupling capacitively to the
islands.  Modeling the offset charges through fluctuating voltage sources that
charge capacitors $C_s$ coupled to the islands yields the relation $n_s^{(j)}=
C_s V_s^{(j)}/2e$ for the stray offset charge $n_s^{(j)}$.  Apart from the
change in rhombus design, the proposed setup remains the same as in
Fig.~\ref{fig:setup}.  Importantly, it is sufficient to add inductive shunts
only to the lower part of the rhombus arms such that no inductive coupling to
the main island is generated.}\label{fig:inductive_shunts}
\end{figure}

The simple rhombus design that we have presented in the main text for clarity
of the discussion suffers from the fact that the charge configuration on the
capacitors of the central islands in the rhombus is strongly susceptible to
fluctuations of stray offset charges that couple capacitively to the islands.
They can be modeled, e.g., through voltage sources $V_s^{(1)/(2)}$ that charge
capacitors $C_s$ coupled to the rhombus islands, yielding $n_s^{(1)/(2)} = C_s
V_{s}^{(1)/(2)}/2e$ for the stray offset charge $n_s^{(1)/(2)}$; see
Fig.~\ref{fig:inductive_shunts}.  These stray offset charges are not
controlled and fluctuate in general independently, destroying the symmetry of
the rhombus arms and lifting the $\pi$-periodicity of the rhombus.  As was
argued theoretically~\cite{doucot:12} and tested
experimentally~\cite{bell:14}, this can be remedied by adding inductive shunts
of strength $L$ to the lower parts of the rhombus arms.  In the classical case
and ignoring the Josephson junctions, adding inductive shunts to the ground in
the central islands of the rhombus reduces this susceptibility by a factor of
$\omega^2/\omega_{LC}^2$, where $\omega$ is the frequency of offset charge
fluctuations and $\omega_{LC} = 1/\sqrt{2 L C_\diamond} = \sqrt{4
E_{C\diamond} E_{L}}/\hbar$ is the plasma frequency of charge oscillations in
the resulting LC resonator with the inductive energy $E_{L} =
(\Phi_0/2\pi)^2/L$.  As was shown in Ref.~\onlinecite{koch:09}, this property
carries over to the quantum case.  As a consequence of the inductive shunts,
the rhombus does not couple any more directly to $n_s^{(1)/(2)}$ but only to
$\dot n_s^{(1)/(2)}$.  The associated noise power changes from the $1/\omega$
form typical for fluctuations of $n_s^{(1)/(2)}$ \cite{astafiev:04} to a much
more benign noise power proportional to $\omega$.

As we have discussed above, the inclusion of inductive shunts reduces the
sensitivity to offset charge fluctuations, but it should of course also
preserve the behavior of the rhombus as an effective $\cos(2\phi)$ element.
This means that in our setup, we cannot make $E_L$ arbitrarily large since our
treatment required that both the band indices and the Bloch momenta $q_i$
associated with the translational symmetry in the variables $\theta_i$ in
absence of inductive shunts remain good variables.  This means that $E_L$
should be a weak perturbation with $E_L \ll E_{J\diamond}, E_{C\diamond}$.  We
can therefore follow the ideas of Refs.~\onlinecite{koch:09,schon:90} and
transform to a basis of Bloch waves.  Solving the $2\pi$-periodic part of the
rhombus perturbatively in $\eta$ and projecting on the lowest band of the
rhombus yields a Hamiltonian of the form
\begin{align}\label{eq:rhombus_effective_hamiltonian_with_inductances}
H_{\diamond}' &= \sum_{j=1,2} \Biggl\{ \frac{E_L}{2} \biggl( i \frac{d}{d q_j}\biggr)^2 
+ f(q_j) \nonumber \\
& \quad
+ g(q_j) \cos(2\phi) + (-1)^{j} h(q_j) \sin(\phi)\Biggr\},
\end{align}
with periodic functions $f$, $g$, $h$ with period $1$.  The absence of a
coupling between $q_1$ and $q_2$ reflects the fact that the phases $\theta_j$
are not coupled by the rhombus Hamiltonian.  The presence of a coupling to
$\cos(2\phi)$ [$\sin(\phi)$] that is even [odd] under the exchange $q_1
\leftrightarrow q_2$ reflects the symmetry under exchange of the rhombus arms
and simultaneous advance of $\phi$ by $\pi$ that we discussed in the main
text.  The additional time-reversal symmetry at half a flux quantum forbids a
coupling to $\cos(\phi)$ or $\sin(2\phi)$ and requires that the functions $f$,
$g$, $h$ possess quadratic expansions around $q=0$ which we find to be of the
form
\begin{align}\label{eq:rhombus_potentials}
f(q)/E_{C\diamond} &= 2 q^2\Bigl(1+ \mathcal{O}\bigl(\eta^2, q^2\bigr)\Bigr) +
\mathcal{O}\bigl(\eta^4\bigr), \nonumber \\
g(q)/E_{C\diamond} &= -\frac{7}{8} \eta^4\Bigl( 1 + \frac{111}{7} q^2 +
\mathcal{O}\bigl(\eta^2, q^4\bigr)\Bigr),
\\
h(q)/E_{C\diamond} &= -8 \eta^2 q^2 \Bigl( 1 + \mathcal{O}\bigl(\eta^2, q^2\bigr)\Bigr).\nonumber
\end{align}
Obviously, for $q_1 = q_2 = 0$ and $E_L = 0$, the Hamiltonian $H_\diamond'$
reproduces the ground state energy $\epsilon_0(\phi)$ that was given in the
main text.  For finite $E_L$, the former rhombus eigenstates with sharp Bloch
momenta $q_1$, $q_2$ are replaced by eigenstates of the Hamiltonian
\eqref{eq:rhombus_effective_hamiltonian_with_inductances}.  Assuming a
symmetric state of the rhombus arms, the ground state energy
$\epsilon_0'(\phi)$ of $H_\diamond'$ can still be expanded in the form
$\epsilon_0'(\phi) = -\sum_n E_{J2n}' \cos(2n\phi)$.  Importantly, $E_L \ll
E_{C\diamond}$ implies $q^2 \sim \sqrt{E_L/4 E_{C\diamond}} \ll 1$ which
guarantees that the term $E_{J2}'$ remains dominant in the expansion of
$\epsilon_0'(\phi)$.  Numerical checks show that this property persists beyond
the perturbative regime in $\eta$ and remains valid also for $\eta = 0.7$.
The inductive shunts thus manage to reduce the susceptibility to charge noise
without spoiling the generation of an effective $\cos(2\phi)$ potential by the
rhombus.

Finally, let us comment on the adiabatic decoupling of the island and the
rhombus for finite $E_L$.  The level spacing of the eigenstates of the
Hamiltonian~\eqref{eq:rhombus_effective_hamiltonian_with_inductances} is of
the order of the plasma energy $\hbar \omega_{LC}$.  In view of the adiabatic
decoupling, it is desirable to have $\hbar \omega_{LC} \gtrsim E_\text{SUSY}$,
i.e., $E_L \gtrsim E_\text{SUSY}^2/E_{C\diamond}$, where $E_\text{SUSY}$ is
the energy scale of the rhombus.  We note that for $\alpha = 2$ and
$\eta=0.7$, we have $E_\text{SUSY} \approx E_{J1} = E_{C\diamond}/10$ such
that choosing $E_L \gtrsim E_\text{SUSY}^2/E_{C\diamond} = E_{C\diamond}/100$
and $E_L \ll E_{C\diamond}, E_{J\diamond}$ is easily possible.  Following
Ref.~\onlinecite{bell:14}, such an inductance can in practice be implemented
through a chain of Josephson junctions.


\end{document}